\documentclass[reprint, prmaterials, amsfonts, amssymb, amsmath, aps, floatfix, article]{revtex4-2}
\usepackage[pdfusetitle]{hyperref}
\usepackage{graphicx}
\usepackage{dcolumn}
\newcolumntype{d}[1]{D{.}{.}{#1}}
\usepackage{bm}
\usepackage{longtable}
\usepackage{mathtools}
\usepackage[normalem]{ulem}

\usepackage{xspace}
\usepackage{xpunctuate}

\hypersetup{
  colorlinks=true,
  linkcolor=blue,
  filecolor=blue,
  citecolor=blue,
  urlcolor=blue,
}

\newcommand{\bro}{Ba$_2$IrO$_4$\xspace}
\newcommand{\sro}{Sr$_2$IrO$_4$\xspace}
\newcommand{\jeff}{J_{\textrm{eff}}}

\usepackage{float}

\begin{document}

\title{Octahedral rotation instability in Ba$_2$IrO$_4$}

\author{Alaska Subedi} 

\affiliation{CPHT, CNRS, \'Ecole polytechnique, Institut Polytechnique
  de Paris, 91120 Palaiseau, France}

\date{\today}

\begin{abstract}
\bro has been refined in the tetragonal $I4/mmm$ phase without
octahedral rotations, and its physical properties have been
interpreted in this high-symmetry structure.  However, the dynamical
stability of this undistorted phase has not previously been
questioned.  It is important to establish whether other lower-symmetry
structures are energetically more favorable because octahedral
rotations control electronic bandwidths and constrain which magnetic
interactions are allowed by symmetry.  Here I compute first-principles
phonon dispersions of $I4/mmm$ \bro including spin-orbit interaction.
I find a nearly-flat nondegenerate unstable branch along the
Brillouin-zone boundary segment $XP$ associated with inplane rotations
of the IrO$_6$ octahedra.  Using group-theoretical analysis, I
enumerate the symmetry-allowed distortions associated with the $X_2^+$
and $P_4$ instabilities and fully relax the resulting structures.
Only five of the twelve possible distortions can be stabilized, and
the energy gain scales with the number of layers that exhibit
octahedral rotations: phases with rotations in every IrO$_6$ layer are
lower by $-5.8$ meV/atom and are nearly degenerate with respect to the
stacking phase.  Electronic structure calculations show that these
rotated phases host a narrow and well-separated half-filled
$\jeff=1/2$ manifold, whereas structures with rotations only in
alternate layers have broader and more entangled bands.  This
motivates a reinvestigation of the crystal structure of \bro and
indicates that octahedral rotations should be considered in modeling
its correlated electronic and magnetic properties.
\end{abstract}

\maketitle

\section{Introduction}

Iridates in perovskite-derived structures with Ir$^{4+}$ ($5d^5$) have
received sustained attention because strong spin-orbit coupling acting
on the $t_{2g}$ manifold can generate a spin-orbit-entangled
half-filled band with dominant $\jeff = 1/2$ character
\cite{Kim2008,Kim2009,Jackeli2009}.  Depending on the crystal
structure and Ir-O-Ir bond angles, the resulting half-filled band can
be sufficiently narrow that a moderate on-site Coulomb repulsion
stabilizes a spin-orbit-assisted Mott state with antiferromagnetism,
inviting comparison with cuprate parent compounds
\cite{Bertinshaw2019}.

Most iridates exhibit octahedral distortions and rotations that are
ubiquitous in perovskites.  Such departures from the ideal octahedral
environment admix the $\jeff = 1/2$ and $\jeff = 3/2$ characters, and
the system is not captured by a strict single-band description.
Furthermore, changes in the Ir-O-Ir bond angles strongly affect the
electronic bandwidths and magnetic superexchange interactions and
anisotropies \cite{Porras2019}.  In this regard, Ba$_2$IrO$_4$ is an
unusual member of this family that has been reported in the tetragonal
$I4/mmm$ structure without octahedral rotations \cite{Okabe2011}.  In
contrast, Sr$_2$IrO$_4$ exhibits staggered rotations that lower the
symmetry \cite{Crawford1994}.  In $I4/mmm$, the inversion center at
the Ir-O-Ir bond midpoint forbids the nearest-neighbor
Dzyaloshinskii-Moriya interaction, which is enabled by rotations in
\sro and underlies its moment canting
\cite{Jackeli2009,Cao1998,Ye2013}.  This higher symmetry in the
absence of rotations makes \bro a natural baseline for isolating how
octahedral rotations modify the $\jeff = 1/2$ electronic structure and
magnetic anisotropies \cite{Moser2014,Katukuri2014}, and it is one
reason it has been compared with the cuprate parent La$_2$CuO$_4$
\cite{Boseggia2013,Nichols2014,Uchida2014}.  Recent work has also
revisited low-energy descriptions and spectroscopic fingerprints of
\bro assuming an undistorted structure
\cite{Clancy2023,Cassol2024,Cassol2025}.  For this assumption to be
meaningful, it is necessary to establish whether the reported $I4/mmm$
phase is dynamically stable.

First-principles calculations comparing \sro and \bro have optimized
the \bro structure starting from the internal coordinates of \sro and
found broadly similar low-energy electronic structures
\cite{Arita2012}.  However, to the best of my knowledge, the dynamical
stability of the $I4/mmm$ \bro has not been systematically tested
using first-principles calculations of the phonon dispersions, nor has the
energy landscape of possible low-symmetry structures been mapped.
Such an investigation would help in identifying energetically
favorable rotations that may be difficult to detect in experiments if
different stackings of the octahedral distortions are nearly
degenerate due to weak interlayer coupling.  Indeed, bulk studies of
\bro have so far been limited to polycrystals \cite{Okabe2011} or tiny
single crystals \cite{Moser2014}.

In this paper, I compute the phonon dispersions of Ba$_2$IrO$_4$ in the
$I4/mmm$ phase to investigate its dynamical stability.  I find a
nearly-flat nondegenerate branch involving inplane rotations of the
IrO$_6$ octahedra that is unstable along the Brillouin zone boundary
line $XP$.  Using group-theoretical analysis, I construct and
relax the symmetry-allowed distortions associated with the unstable
modes at $X$ and $P$.  I find that the energy gain scales with
the number of layers that exhibit octahedral rotations.  The
lowest-energy phases differ mainly by the stacking of the octahedral
rotations and are degenerate within numerical accuracy. Band structure
calculations show that these distortions cause similar narrowing and
splitting of the $\jeff = 1/2$ bands, making it difficult to infer
stacking from inplane electronic structure alone.

\section{Computational Approach}

The phonon calculations presented here were performed using density
functional perturbation theory as implemented in the {\sc quantum
  espresso} plane-wave code version 7.2 \cite{qe}.  I used the version
1.0.0 of the fully-relativistic pseudopotential collection generated
by Dal Corso \cite{pslib}, with the valence band configurations Ba
$5s^2 6s^2 5p^6$, Ir $6s^2 6p^0 5d^7$, and O $2s^2 2p^4$.  The
calculations were done within the generalized gradient approximation
of Perdew, Burke, and Ernzerhof \cite{pbe}.  The cutoffs for the
basis-set and charge-density expansions were set to 60 and 600 Ry,
respectively.  An $8\times8\times8$ $k$-point grid was used for the
Brillouin zone integration with a Marzari-Vanderbilt smearing of 0.01
Ry.  The non-spin-polarized dynamical matrices were calculated
including the spin-orbit interaction on a $4\times4\times4$ $q$-point
grid, and the phonon dispersions were obtained using Fourier
interpolation.  {\sc amplimodes}, {\sc findsym}, and {\sc spglib} were
used in the symmetry analysis of the unstable phonons and distorted
structures \cite{ampli,findsym,spglib}.  I used the {\sc isotropy}
code to determine the space group and order parameter directions of
the low-symmetry structures due to the unstable modes \cite{isotropy}
and generated them on commensurate supercells that were then reduced
to 56-atom structures.

Structural relaxations including the spin-orbit interaction of such
large structures is a computationally intensive endeavor.  Therefore,
the distorted structures were relaxed using the {\sc vasp} package
version 6.4.2 \cite{vasp}.  They were done using the projector
augmented-wave pseudopotentials with the valence configurations of Ba
$5s^2 5p^6 6s^{1.99} 5d^{0.01}$, Ir $6s^1 5d^8$, and O $2s^2 2p^4$.  The
Brillouin zone integration was done using a $12\times12\times3$
$k$-point grid.  A relatively high 800 eV basis-set cutoff was used
because the relaxation calculations also included the spin-orbit
interaction.

\section{Results and Discussion}

\begin{figure}
  \includegraphics[width=\columnwidth]{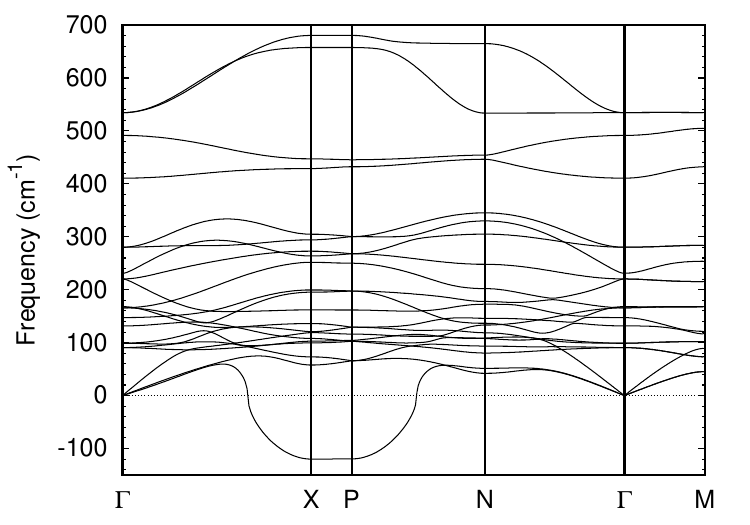}
  \caption{Calculated non-spin-polarized phonon dispersions of
    fully-relaxed \bro in the $I4/mmm$ phase obtained using the PBE
    functional and including the spin-orbit interaction.  The
    high-symmetry points are $\Gamma$ $(0,0,0)$, $X$ $(1/2,1/2,0)$,
    $P$ $(1/4,1/4,1/4)$, $N$ $(0,1/2,0)$, and $M$ $(1/2,1/2,1/2)$ in
    terms of the primitive body-centered tetragonal reciprocal basis
    vectors proportional to $(0,1,1)$, $(1,0,1)$, and $(1,1,0)$.
    Imaginary frequencies are indicated by negative values.}
  \label{fig:ph}
\end{figure}

\begin{figure}
  \includegraphics[width=0.45\columnwidth]{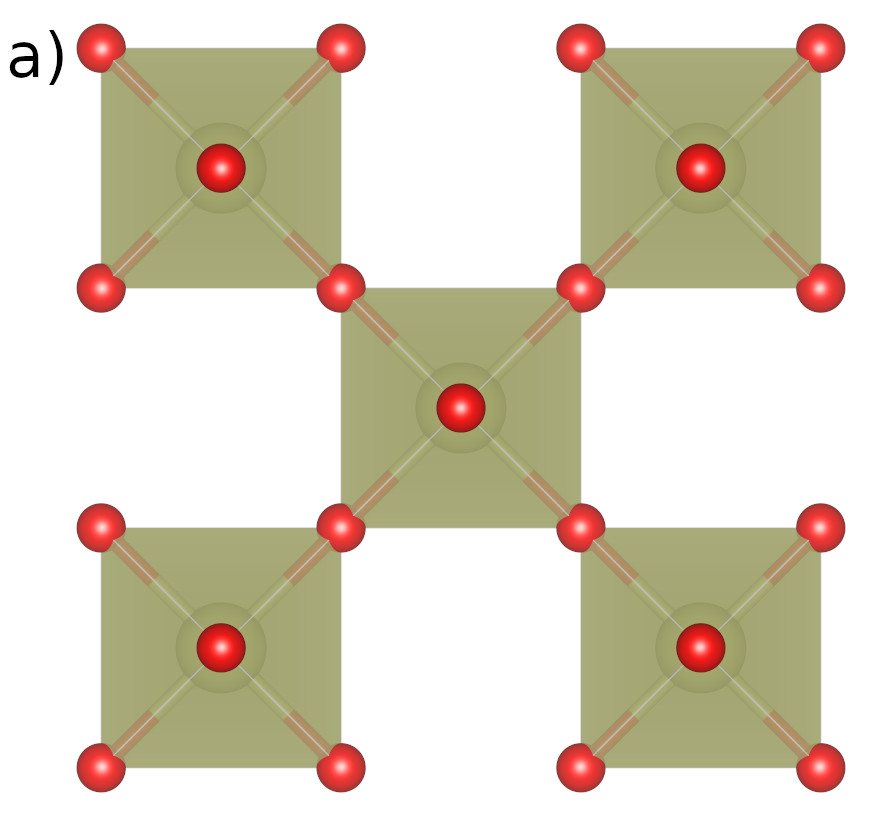}
  \quad
  \includegraphics[width=0.45\columnwidth]{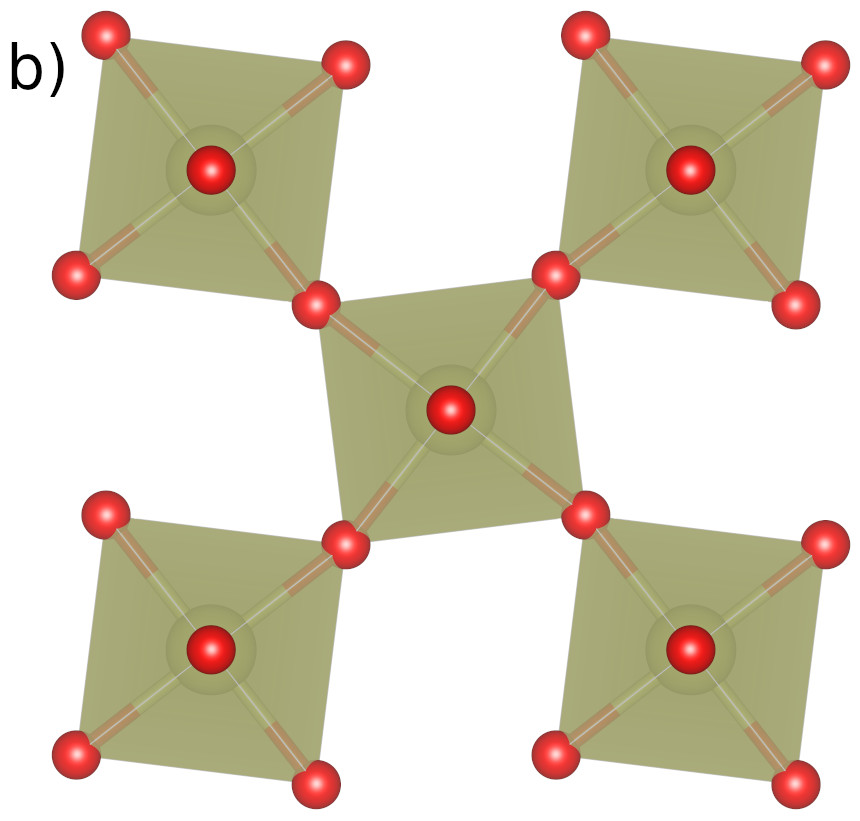}
  \caption{a) Octahedral layer in the parent $I4/mmm$ phase.  b)
    Octahedral rotation due to the unstable phonon branch along $XP$.
    Only the basal oxygen ions move due to the instabilities.  The
    octahedral rotation patterns at $X$ and $P$ are identical within
    the plane; they differ only in the phase of the rotation along the
    out-of-plane direction.  The oxygen ions are denoted by red
    spheres.  The iridium ions reside inside the octahedra.}
  \label{fig:octa}
\end{figure}

The calculated phonon dispersions of non-spin-polarized \bro in the
currently accepted $I4/mmm$ structure are shown in Fig.~\ref{fig:ph}.
They reveal a nondegenerate branch that is unstable along the segment
$XP$ of the Brillouin zone, which indicates that the $I4/mmm$ phase is
dynamically unstable.  The unstable branch is nearly flat, with the
calculated imaginary frequencies of 119.9$i$ and 119.7$i$ cm$^{-1}$ at
$X$ and $P$, respectively.  The eigenvectors of these modes are also
identical and involve only the displacements of the O ions lying in
the octahedral plane such that the octahedra are rotated within the
plane, as shown in Fig.~\ref{fig:octa}.  Hence, the distorted
structures due to these instabilities differ primarily by how the
inplane octahedral rotations propagate along the out-of-plane
direction.

\begin{figure}
  \includegraphics[width=\columnwidth]{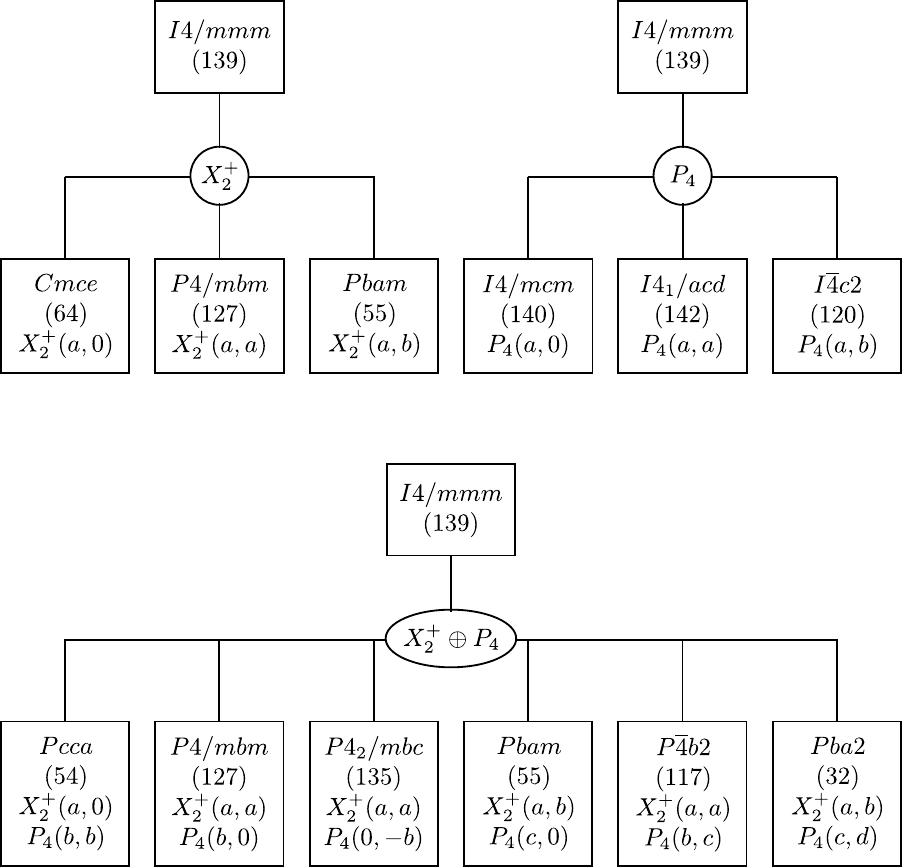}
  \caption{Isotropy subgroups and the corresponding order parameters
    of the $X_2^+$ and $P_4$ irreps of the $I4/mmm$ space group. The
    space group numbers are given in parentheses.}
  \label{fig:iso}
\end{figure}

Both $X$ and $P$ points have two elements in their respective stars
$\{(1/2, 1/2, 0)$, $(0, 0, 1/2)\}$ and $\{(1/4, 1/4, 1/4)$,
$(3/4,3/4,3/4)\}$. Therefore, the order parameter subspace spanned by
the unstable phonon modes at $X$ and $P$ is four dimensional even
though the modes are nondegenerate.  I find the instabilities at $X$
and $P$ to have the irreducible representation (irrep) $X_2^+$ and
$P_4$, respectively.  Group-theoretical analysis using the {\sc
  isotropy} code shows that there are twelve distinct distorted
structures that can arise out of these instabilities, and they are
shown in Fig.~\ref{fig:iso}.  I generated these structures using the
calculated eigenvectors of the unstable phonon modes and then fully
relaxed them by minimizing both the atomic forces and lattice stresses
without the presence of any magnetic order.

\begin{table}
    \caption{\label{tab:ene} Total energies of the five distorted
      structures due to the $X_2^+$ and $P_4$ instabilities present in
      the $I4/mmm$ phase of Ba$_2$IrO$_4$ that could be stabilized
      after structural relaxations.  These were obtained using
      non-spin-polarized calculations that included spin-orbit
      interaction.}

    \begin{ruledtabular}
      \begin{tabular}{l l l d{3.2}}
        Space group     & $X_2^+$ & $P_4$ & \multicolumn{1}{c}{Energy}
        (meV/atom)\\
        \hline
        $I4/mmm$   &         &          &  0.0 \\
        $P4/mbm$   & $(a,a)$ &          & -2.5 \\
        $I4/mcm$   &         & $(a,0)$  & -2.5 \\
        $Cmce$     & $(a,0)$ &          & -5.8 \\
        $I4_1/acd$ &         & $(a,a)$  & -5.8 \\
        $P4_2/mbc$ & $(a,a)$ & $(0,-b)$ & -5.8 \\
      \end{tabular}
    \end{ruledtabular}
\end{table}

Only five out of the twelve possible distortions could be stabilized
after structural relaxations, and their relative energies with respect
to the undistorted parent structure are given in Table~\ref{tab:ene}.
The low-symmetry structures are grouped into two values of energy gain
within the numerical precision.  Two structures show an energy gain of
$-2.5$ meV/atom, while three have a gain of $-5.8$ meV/atom.  While
considering the magnitude of these energy gains, it is worthwhile to note
that only the basal O ions move due to the instabilities.  Taking into
account that there are two basal O ions per formula unit, the energy
gains are $-8.8$ and $-20.3$ meV per basal O, which are substantial.

\begin{figure}
  \includegraphics[width=\columnwidth]{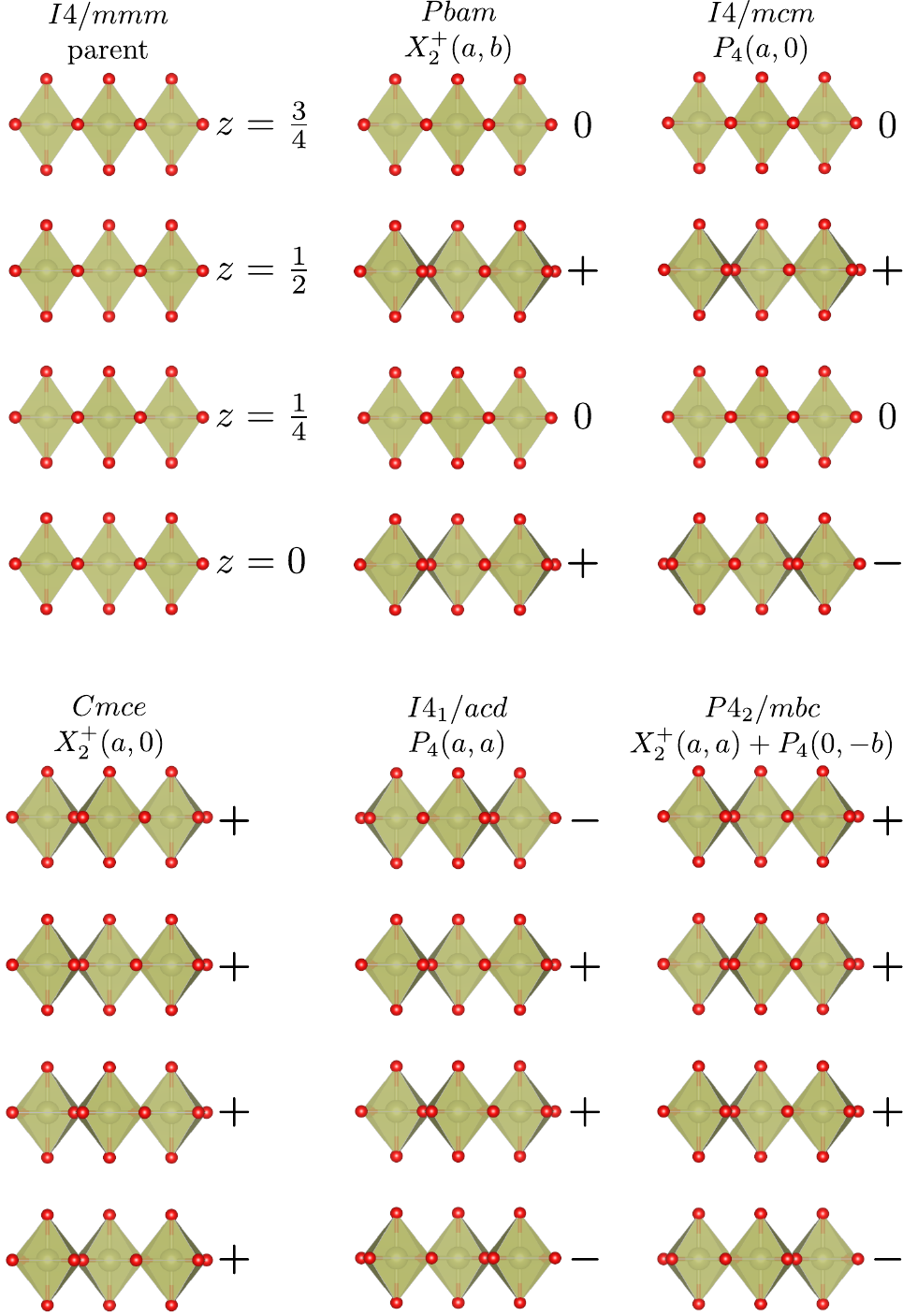}
  \caption{Rotation patterns of the IrO$_6$ octahedral layers in the
    structures that could be stabilized after structural relaxation
    calculations. The relative patterns of the rotations between
    different layers are denoted by $+$ and $-$. The layers without
    octahedral rotation are denoted by $0$. Note that the $z = \{0,
    \frac{1}{2}\}$ and $z=\{\frac{1}{4},\frac{3}{4}\}$ layers are
    shifted with respect to each other by an inplane wavevector of
    $(\frac{1}{2},0)$.}
  \label{fig:rot}
\end{figure}

Interestingly, the volume of the IrO$_6$ octahedra is $\sim$11.7
\AA$^3$ in all the relaxed structures, and the differences in the
relative energies seem to derive only from the energy gain due to
octahedral rotations.  The rotational patterns of the octahedra in all
the five structures that could be stabilized are shown in
Fig.~\ref{fig:rot}.  In the $P4/mbm$ $X_2^+(a,a)$ and $I4/mcm$ $P_4(a,0)$
structures that show an energy gain of $-2.5$ meV/atom, the IrO$_6$
octahedra are rotated in only the alternate layers $z = 0$ and
$1/2$.  The Ir-O-Ir angle in the rotated layers is 166$^\circ$
in these structures.  Between these layers with rotations, the layers
at $z = 1/4$ and $3/4$ occur without octahedral
rotations in both the structures.  The two structures differ only by
the phase of the octahedral rotations in the out-of-plane direction.
The rotations are in phase in the $P4/mbm$ structure, whereas they are
out of phase in the $I4/mcm$ structure.

Meanwhile, the octahedra in every layer are rotated to the Ir-O-Ir
angle of 164$^\circ$ in the three structures $Cmce$ $X_2^+(a,0)$,
$I4_1/acd$ $P_4(a,a)$, and $P4_2/mbc$ $X_2^+(a,a) + P_4(0,-b)$ that
exhibit the largest energy gain of $-5.8$ meV/atom, these structures
again differing only by the phase of the octahedral rotations in the
out-of-plane direction.  In the $Cmce$ structure, the rotations are in
phase in the $z = \{0, 1/2\}$ and $z = \{1/4, 3/4\}$ pairs of layers.
The rotations are out of phase in both the pairs in the $I4_1/acd$
structure.  Finally, the rotations are out of phase in the $z = \{0,
1/2\}$ pair of layers and in phase in the $z = \{1/4, 3/4\}$ pair in
the $P4_2/mbc$ structure.  It is remarkable that the energy gain in
these structures depends almost only on the number of layers with
octahedral rotations.  As discussed above, the insensitivity of the
total energy to the phase of the interlayer octahedral rotations
microscopically derives from the flatness of the unstable branch along
$XP$.

\begin{figure}[!htb]
  \includegraphics[width=\columnwidth]{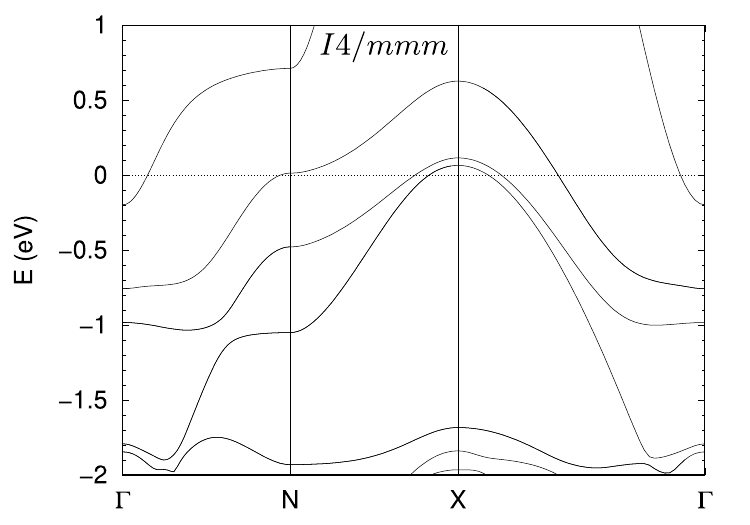}
  \includegraphics[width=\columnwidth]{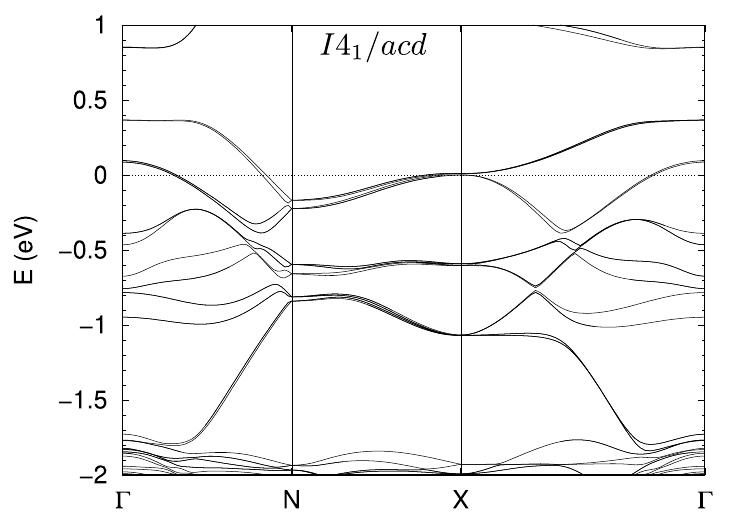}
  \includegraphics[width=\columnwidth]{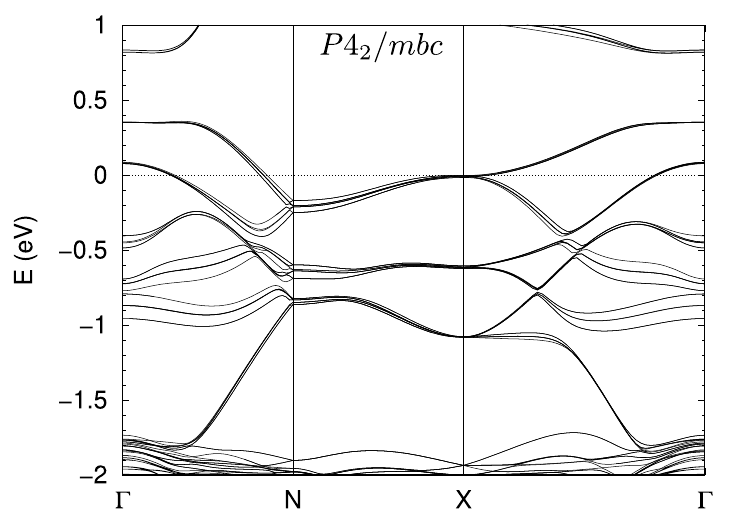}
  \caption{Calculated band structures of \bro in the (top) $I4/mmm$,
    (middle) $I4_1/acd$, and (bottom) $P4_2/mbc$ phases obtained using
    the PBE functional and including the spin-orbit interaction. The
    two low-symmetry structures involve octahedral rotations in all
    layers, and their $\jeff = 1/2$ manifold is narrow, well
    separated, and half filled.}
  \label{fig:bands}
\end{figure}

The Ir-O-Ir angle in the sister compound \sro is $\sim$157$^\circ$.
Previous calculations have shown that the octahedral rotations in \sro
narrow the effective width of the $\jeff = 1/2$ manifold, which lowers
the critical $U$ needed to obtain the insulating paramagnetic state
relative to the undistorted phase in DMFT calculations
\cite{Martins2011}.  To find out if octahedral rotations cause a
similar effect in the electronic structure of \bro, I studied the
effect of octahedral rotations on its electronic structure.

Fig.~\ref{fig:bands} shows the non-spin-polarized band structures
including spin-orbit interaction of the undistorted $I4/mmm$ and
distorted $I4_1/acd$ and $P4_2/mbc$ structures that have the largest
energy gain of $-5.8$ meV/atom.  The band structures of \bro in the
$I4/mmm$ and $I4_1/acd$ phases are similar to those previously
obtained for \sro in the respective structures
\cite{Uchida2014,Cassol2024,Martins2011,Arita2012}.  In the parent
$I4/mmm$ phase, the $\jeff = 1/2$ band extends from $-0.8$ to $0.6$ eV
and is half-filled \cite{Uchida2014,Cassol2024}.  It is well separated
from the valence $\jeff = 3/2$ and conduction $d_{x^2-y^2}$ bands,
which briefly cross the Fermi level at $X$ and $\Gamma$, respectively.
As discussed before, octahedral rotations mix the $\jeff = 1/2$ and
$3/2$ characters.  However, a half-filled manifold with dominant
$\jeff = 1/2$ character that is separated from other bands also occurs
in the $I4_1/acd$ phase of \sro \cite{Kim2008,Martins2011,Arita2012}.
\bro in this phase exhibits similar electronic structure
\cite{Arita2012}.  The half-filled $\jeff = 1/2$ manifold narrows and
is now situated between $-0.4$ and $0.4$ eV, which is a 0.6 eV
reduction in the band width relative to the undistorted phase.  It is
the only manifold to cross the Fermi level and is separated from both
the $\jeff = 3/2$ and $e_g$ bands that lie below and above it,
respectively.
 
The band structure of the $P4_2/mbc$ phase shown in the bottom panel
of Fig.~\ref{fig:bands} is remarkably similar to that of the
$I4_1/acd$ phase, differing almost only in the doubling of the number
of bands due to the bigger unit cell.  It again features a narrow
half-filled $\jeff = 1/2$ manifold that is separated from others.  The
bands that appear extra compared to the $I4_1/acd$ phase almost
overlap.  This similarity is not unreasonable considering the large
separation of the octahedral layers that leads to small splitting due
to suppressed out-of-plane hopping. But it is striking that phase of
the octahedral rotation has a minimal impact on the electronic
structure.  This implies that it will be difficult to distinguish the
two phases by measuring the inplane electronic structure alone.

\begin{figure}
  \includegraphics[width=\columnwidth]{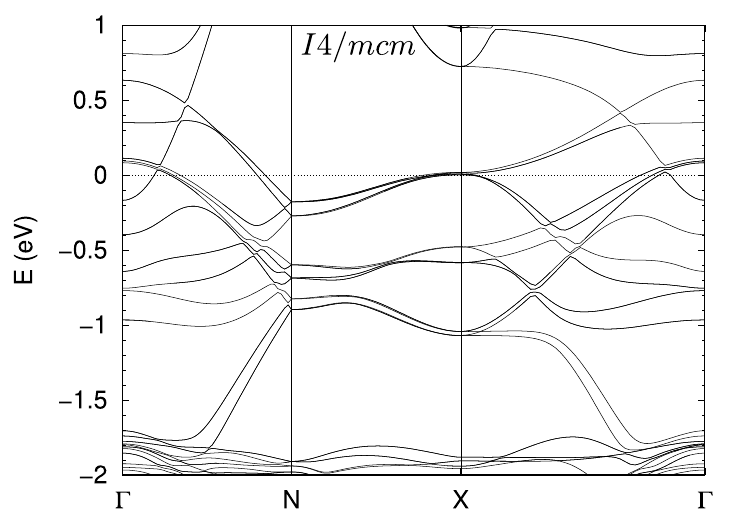}
  \caption{Calculated PBE band structure incorporating spin-orbit
    interaction of \bro in the $I4/mcm$ phase that involves octahedral
    rotations in only the alternate layers.  The $\jeff = 1/2$ bands
    lying around the Fermi level are broader and cross with other
    bands.}
  \label{fig:bandP4a0}
\end{figure}

The band structure of the $I4/mcm$ phase shown in
Fig.~\ref{fig:bandP4a0} incorporates qualitative features from both
the undistorted $I4/mmm$ and distorted $I4_1/acd$ phases. A $\jeff =
1/2$ manifold similar to the one found in the $I4_1/acd$ structure
does form in the $I4/mcm$ phase.  However, this manifold is not
separated from the $\jeff = 3/2$ and $e_g$ bands which both cross the
Fermi level near $\Gamma$ of the now smaller Brillouin zone.
Furthermore, the bands from all these three manifolds have a larger
band width.  This indicates that octahedral rotations not only narrow
the bands, but also cause band hybridzations that separate the three
manifolds. Importantly, this shows that having octahedral rotations in
all the IrO$_6$ layers is necessary to get a separated and narrow
$\jeff = 1/2$ manifold that is half filled.

Lastly, it is worth noting that the results presented here do not take
into account the anharmonic and correlation effects that are needed to
fully describe the physics of this material.  However, they do
motivate a careful reinvestigation of the crystal and electronic
structures in large single crystals.  It is remarkable that the
calculated widths of the $\jeff = 1/2$ bands in the undistorted
$I4/mmm$ phase of \bro and \sro, with values of $\sim$1.4 and
$\sim$1.7, respectively, differ by a modest value.  Also noteworthy is
the comparable widths of the $\jeff = 1/2$ manifold in the $I4_1/acd$
phases, which are $\sim$0.8 and $\sim$1.0 eV in \bro and \sro,
respectively.  DMFT calculations show that the $I4/mmm$ and $I4_1/acd$
phases of \sro become insulating around onsite Coulomb $U$ of 3.0 and
2.1 eV, respectively \cite{Martins2011}.  The difference in $U$ is
much larger than the differences in the widths of the $\jeff = 1/2$
bands in the corresponding phases of \bro and \sro. This suggests that
although the bands in \bro are slightly narrower, octahedral
distortions might still be necessary to induce the Mott insulating
phase observed in this material.

\section{Summary and Conclusions}

In summary, I have investigated the structural instabilities of
$I4/mmm$ \bro using first-principles phonon dispersions and structural
relaxation calculations.  I find a nearly-flat nondegenerate branch
that is unstable along the Brillouin zone boundary segment $XP$, with
imaginary frequencies of about $120i$ cm$^{-1}$ at both $X$ and $P$.
The instability involves inplane rotations of the IrO$_6$ octahedra.
I used group-theoretical analysis to enumerate the symmetry-allowed
distortions associated with the unstable phonon modes at $X$ and $P$
and fully relaxed them to obtain the relative energies.  Only five out
of twelve possible low-symmetry structures are lower in energy after
relaxation.

Among these, $P4/mbm$ and $I4/mcm$ show an energy difference of $-2.5$
meV/atom relative to $I4/mmm$, while $Cmce$, $I4_1/acd$, and
$P4_2/mbc$ are degenerate within numerical accuracy at $-5.8$
meV/atom.  The energy lowering saturates once octahedral rotations
occur in all IrO$_6$ layers.  This can be realized by the $X_2^+$
distortion, the $P_4$ distortion, or their combination, yielding
$Cmce$, $I4_1/acd$, and $P4_2/mbc$, respectively.  Condensing both
instabilities does not produce an additional energy lowering within
numerical accuracy, and the three phases differ only in the stacking
of the octahedral rotations.  The dependence of the energetics
primarily on the number of IrO$_6$ layers that exhibit octahedral
rotations is consistent with the flatness of the unstable branch along
$XP$ and the layered nature of the material.  This also suggests that
longer-period stacking variants may be competitive.  The near
degeneracy of distinct stacking sequences suggests that stacking
faults or short-range stacking correlations may occur, which could be
probed by diffuse scattering in x-ray or neutron diffraction, or by
local structural probes such as pair distribution function analysis or
scanning transmission electron microscopy-based imaging.

Electronic structure calculations show that when octahedral rotations
occur in all IrO$_6$ layers the $\jeff = 1/2$ band is narrowed and
cleanly separated, while the inplane dispersions are essentially
independent of the stacking periodicity.  Therefore, it will be
difficult to distinguish the lowest-energy structures using purely
inplane electronic probes alone.


\begin{acknowledgments}
I am grateful to B.J.\ Kim for discussions that motivated this work.
This study was provided with high-performance computing resources by
GENCI at TGCC thanks to the grant 2025-A0190913028 on the Rome
partition of the supercomputer Joliot-Curie.
\end{acknowledgments}

\bibliography{ba2iro4-struct-paper}

\end{document}